\documentclass[aps, prl,twocolumn, showpacs, groupedaddress]{revtex4}  % for review and submission 

\usepackage{graphicx}    % needed for figures
\usepackage{dcolumn}    % needed for some tables
\usepackage{bm}             % for math
\usepackage{amssymb}   % for math

\begin{document}
\newcommand{\g}{{\gamma}}  
\newcommand{\Dzero}{D0\ } 
\newcommand{\Pt}{{p_T}}   
\newcommand{\e}{{\epsilon}}   

%\leftline{Version 3.6 as of Mar 9, 2008} 
%\leftline{Primary authors: O. Atramentov, Y. Gershtein, A. Melnitchouk}
%\rightline{To be submitted to PRL}
%\rightline{Comments to authors by Mar 4}

% the following line is for submission, including submission to the arXiv!!
\hspace{5.2in} \mbox{FERMILAB-PUB-08/057-E}

%\title{Search for a light Higgs boson in the $\gamma\gamma+X$ final state with the D0 detector at $\sqrt{s}=1.96$~TeV}
\title{Search for $h_f\rightarrow\gamma\gamma$ with the D0 detector at $\sqrt{s}=1.96$~TeV}

% LIST_OF_AUTHORS_R2.TEX               2/19/08              
%
\author{V.M.~Abazov$^{36}$}
\author{B.~Abbott$^{75}$}
\author{M.~Abolins$^{65}$}
\author{B.S.~Acharya$^{29}$}
\author{M.~Adams$^{51}$}
\author{T.~Adams$^{49}$}
\author{E.~Aguilo$^{6}$}
\author{S.H.~Ahn$^{31}$}
\author{M.~Ahsan$^{59}$}
\author{G.D.~Alexeev$^{36}$}
% ...

\affiliation{\vspace{0.1 in}(The D\O\ Collaboration)\vspace{0.1 in}}
% ...
\affiliation{$^{6}$University of Alberta, Edmonton, Alberta, Canada,
                Simon Fraser University, Burnaby, British Columbia, Canada,
                York University, Toronto, Ontario, Canada, and
                McGill University, Montreal, Quebec, Canada}
% ...
\affiliation{$^{29}$Tata Institute of Fundamental Research, Mumbai, India}
% ...
\affiliation{$^{31}$Korea Detector Laboratory, Korea University, Seoul, Korea}
% ...
\affiliation{$^{36}$Joint Institute for Nuclear Research, Dubna, Russia}
% ...
\affiliation{$^{49}$Florida State University, Tallahassee, Florida 32306, USA}
% ...
\affiliation{$^{51}$University of Illinois at Chicago,
                Chicago, Illinois 60607, USA}
% ...
\affiliation{$^{59}$Kansas State University, Manhattan, Kansas 66506, USA}
% ...
\affiliation{$^{65}$Michigan State University,
                East Lansing, Michigan 48824, USA}
% ...
\affiliation{$^{75}$University of Oklahoma, Norman, Oklahoma 73019, USA}
% ...

\date{March 10, 2008}

\begin{abstract}
We report the results of a search for a narrow resonance decaying into two photons in 1.1 fb$^{-1}$ of data collected by the \Dzero experiment at the Fermilab Tevatron Collider during the period 2002--2006. We find no evidence for such a resonance and set a lower limit on the mass of a fermiophobic Higgs boson of $m_{h_f}>100$ GeV at the 95\%~C.L. This exclusion limit exceeds those obtained in previous searches at the Tevatron and covers a significant region of the parameter space $B(h_{f}\to\gamma\gamma)$ vs. $m_{h_f}$ which was not accessible at the CERN LEP Collider.

\end{abstract}

\maketitle 
In the standard model (SM), the Higgs field is responsible for both electroweak symmetry breaking and generating elementary fermion masses. And while the SM describes our world at current experimentally accessible energies, the exact mechanism for electroweak symmetry breaking remains a mystery. 

Di-photon decays of the Higgs boson are suppressed at tree level, and in the SM such decays have a very small branching fraction, $10^{-3}-10^{-4}$. However, in a more general framework where the parameter content of the theory is richer, such decays can be enhanced.  
In the situation where the Higgs--fermion couplings are substantially suppressed, the full decay width of the Higgs boson would be shared mostly between the $WW$, $ZZ$, and $\gamma\gamma$ decay modes. Such a scenario,  the so-called ``fermiophobic" Higgs boson, arise in a variety of models, e.g. \cite{landsberg_cite,landsberg_cite2, landsberg_cite3}. In all these cases, for masses $m_h<100$ GeV, the Higgs boson dominantly decays to photon pairs. 

Experimental searches for fermiophobic Higgs bosons ($h_f$) at the CERN LEP Collider and the Fermilab Tevatron Collider have yielded negative results. Mass limits have been set in a benchmark model that assumes that the coupling $h_fVV$ ($V\equiv W^\pm, Z$) has the same strength as in the SM and that all fermion branching ratios (B) are exactly zero. Combination of results obtained by the LEP collaborations \cite{aleph,delphi, l3,opal} using the process $e^+e^-\rightarrow h_fZ,~h_f\rightarrow \g\g$ yielded the lower bound $m_h>109.7$ GeV at the 95\%~C.L. \cite{LEPlimit}.  In Run I of the Tevatron, lower limits on $m_{h_f}$ from the \Dzero and CDF collaborations are respectively 78.5 GeV ~\cite{d0limit} and 82 GeV ~\cite{cdflimit}, using the processes $q\bar{q}'\rightarrow V^*\rightarrow h_fV,~h_f\rightarrow \g\g$, with the dominant contribution coming from $V = W^\pm$. 
  
In this Letter we perform a search for the inclusive production of di-photon final 
states via the Higgsstrahlung and vector boson fusion processes: $p\bar{p}\rightarrow h_fV \rightarrow \gamma\gamma+X$ and $p\bar{p}\rightarrow VV\rightarrow h_f\rightarrow \gamma\gamma+X$, respectively.  The total integrated luminosity of the data used for this search is 1.10 $\pm$ 0.07 fb$^{-1}$. 

The \Dzero detector comprises a central tracking system in a 2 T superconducting solenoid, a liquid-argon/uranium sampling calorimeter, and a muon spectrometer. The calorimeter consists of a central section (CC) covering the pseudorapidity range $|\eta|<1.1$, which is defined as $\eta\equiv -\log [\tan(\frac{\theta}{2})]$ where $\theta$ is the polar angle with respect to the proton beam direction, and two endcaps (EC) extending coverage to $|\eta|<4.2$, each housed in a separate cryostat.  The electromagnetic (EM) section of the calorimeter has four layers with longitudinal  depths of 2$X_0$, 2$X_0$, 7$X_0$, and 10$X_0$ that provide full containment of EM particles (photons and electrons).  The calorimeter layers have transverse segmentation of $\delta\phi\times\delta\eta=0.1\times0.1$ (where $\phi$ is the azimuthal angle), except in the third layer, where it is $0.05\times0.05$, which allows for accurate determination of the position of EM particles. Immediately before the inner layer of the central EM calorimeter there is a central preshower detector (CPS) formed of 2$X_0$ of absorber followed by several layers of scintillating strips with embedded wavelength-shifting fibers. A complete description of the \Dzero detector can be found in \cite{d0det}.

We select events that satisfy single EM triggers which become fully efficient for EM showers with transverse momentum $\Pt >30$~GeV. Photons and electrons are identified in two steps: the selection of EM clusters, and their subsequent separation into those caused by photons and those caused by electrons. EM clusters are selected from calorimeter clusters by requiring that (i) at least 97\% of the 
energy be deposited in the EM section of the calorimeter, (ii) the calorimeter isolation be less than 0.07 (isolation is defined as $[E_{\text {tot}}(0.4) - E_{\text{EM}}(0.2)]/E_{\text{EM}} (0.2)$, where $E_{\text {tot}}(0.4)$ is the total shower energy in a cone of radius $R=\sqrt{(\Delta \eta)^2+(\Delta \phi)^2} = 0.4$, and $E_{\text{EM}}(0.2)$ is the EM energy in a cone with $R = 0.2$), (iii) the transverse, energy-weighted shower width be less than 0.04 rad (i.e. consistent with an EM shower profile), and (iv) the scalar $\Pt$ sum of all tracks originating from the 
primary vertex in an annulus of $0.05 < R < 0.4$ around the cluster be less than 2 GeV. The cluster is then defined as an electron if there is a reconstructed track (or electron-like pattern of hits in the tracker) associated with it and a photon otherwise. We also consider EM jets (jets with a leading $\pi^0$ or $\eta$) defined as EM clusters that pass all cuts required for photon candidates except the track isolation requirement. We will refer to them as ``{\sl j}" or ``jet". We select events that have at least two photons in the central calorimeter ($|\eta|<1.1$) with transverse momenta $\Pt>25$ GeV. Events are required to have the primary vertex close to the beam axis and within 60 cm of the geometrical center of the detector. Identification of the primary vertex in the event is  important, as it affects the calculation of the $\Pt$ of a photon candidate and its track isolation. Despite the fact that photons do not leave tracks, the probability to reconstruct a primary vertex is high, 99.5\%, due to the underlying event activity.

The Higgs boson produced in the models considered has higher transverse momentum $q_T^{\gamma\gamma}$ than that of the two-photon system of the backgrounds. Therefore, we select events with $q_T^{\gamma\gamma}>$ 35 GeV. For simplicity, we choose a fixed cut value which is below the optimal cut value for  Higgs boson masses starting from 70 GeV.  After all selection criteria, we are left with 196 (1509) di-photon events with $q_T^{\g\g}>35$ ($q_T^{\g\g}<35$) GeV for invariant masses above 65~GeV.

The dominant background comes from direct di-photon production (DDP) processes. The other major background comes from events in which jets are misidentified as photons: $\g j$ processes where a quark or a gluon fragmented into an energetic $\pi^0$ or $\eta$ and is reconstructed as a photon, and the multijet background where two jets are mis-identified as photons. 

Another source of di-photon background comes from events in which electrons are misidentified as photons: the decay of a $Z$ boson where electrons are reconstructed as photons if there are no associated tracks, and processes with one real electron coming from the decay of a $W^\pm$ boson produced in association with a real photon or a jet misreconstructed as a photon. The veto of electron-like patterns of hits in the tracker reduces electron backgrounds by a factor of five, while keeping the photon efficiency high. We measure that ($91\pm3$)\% of photon candidates in $Z/\g^*\rightarrow e^+ e^- \g$ data satisfy the anti-track activity requirement. The contribution of events with one or two real electrons is obtained by applying the probability for an electron to fail the track requirement and be reconstructed as a photon ($1.5^{+3.0}_{-1.5}$\%) to the $Z$ boson, Drell-Yan, and $W^\pm+X$ event yields. This background is estimated to be less than one event.

We estimate the relative contributions of the $\g\g$, $\g j$, and $jj$ backgrounds using the difference in the energy weighted width of the energy deposition in the CPS, $\sigma_E^{\text{CPS}}$. The width is generally narrower for photons than for jets. We construct one-dimensional templates as a function  of $x=\sigma_E^{\text{CPS}}$ for photons [$G(x)$] and jets  [$J(x)$]. The $G(x)$ is constructed using radiative  $Z/\g^*\rightarrow \ell^+\ell^-\gamma $ $(\ell=e,\mu)$ decays in data and the $J(x)$ is taken from the $jj$ data sample. From these we construct two-dimensional profiles for the three components $\g\g$, $\g j$, and $jj$, as follows: $GG(x,y) = G(x) \cdot G(y)$, $GJ(x,y) =  0.5\cdot[G(x)\cdot J(y)+J(x)\cdot G(y)]$, and $JJ(x,y)  = J(x)\cdot J(y)$. Further, using these two-dimensional templates we construct a fitting function: $c_0\cdot[GG(x,y) + c_1\cdot JJ(x,y)+c_2\cdot GJ(x,y)]$. The parameters are chosen so that $c_0$ is equal to the number of $\g\g$ events and responsible for the overall normalization, and $c_1$ and $c_2$ determine the contributions of $jj$ and $\g j$ events relative to $\g\g$. 

For the di-photon candidate data sample, we make a two-dimensional distribution of $\sigma_E^{\text{CPS}}$. For each event we randomly decide whether the leading photon is plotted along the $x$- or the $y$-axis. We fit this distribution with the function defined above to determine the individual components: $c_0 =  131\pm 22\pm7~\text{events}$, $c_1 =  0.35\pm0.19\pm0.06$, and $c_2 =  0.13\pm0.28\pm0.13$, where the first error is the statistical error of the fit, and the second is the systematic uncertainty obtained from variations of the fitting range, binning of the templates, and the source of the photon template. 

\begin{figure}[htp]
\begin{center}
\includegraphics[scale=0.45, angle=90]{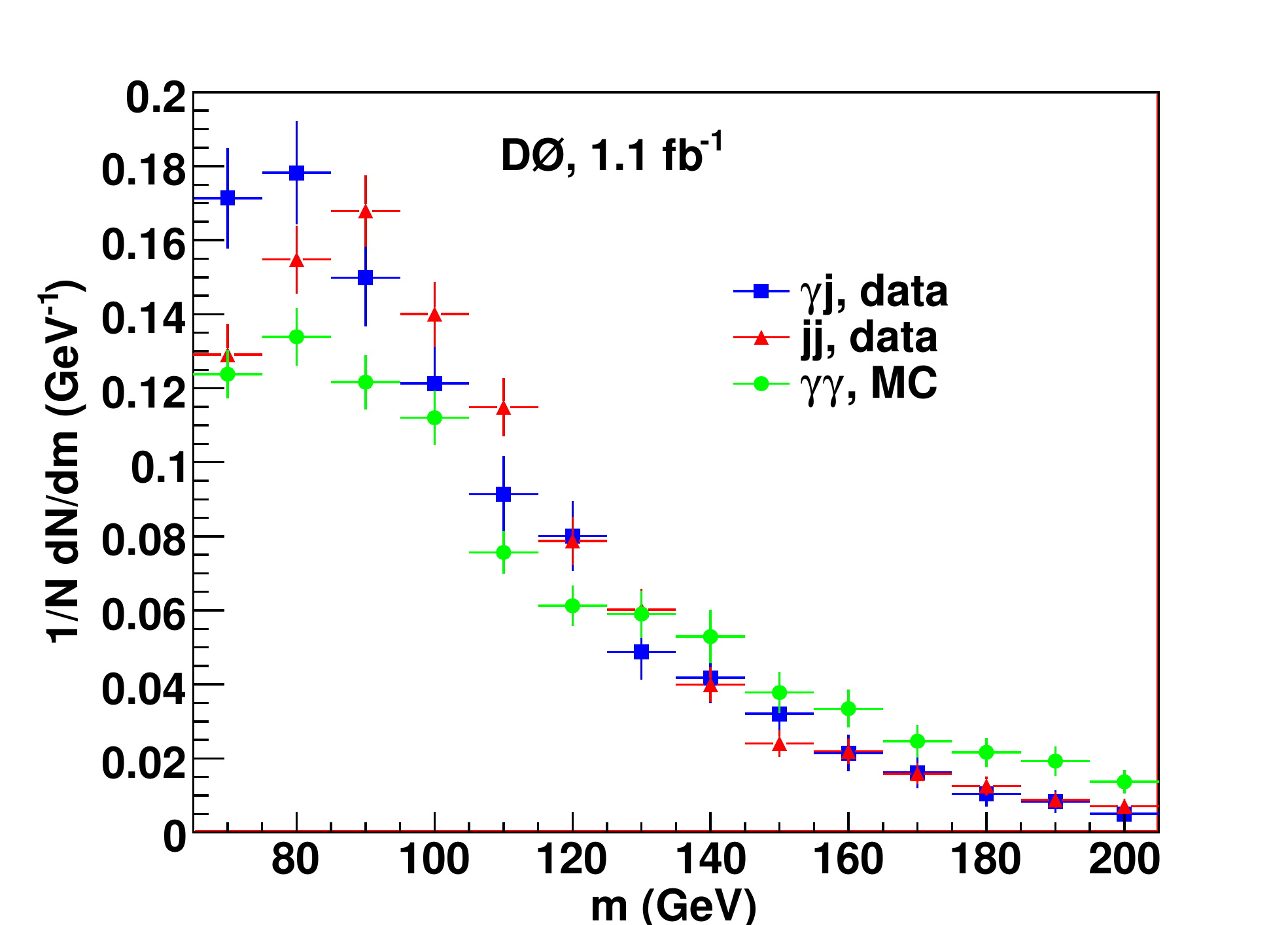}
\end{center}
\caption{Normalized distributions of the invariant mass, $m$, of $\g\g$ (circles), $\g j$ (squares), and $jj$ (triangles).\label{fig:fig_1}}
\end{figure}

The next step is to use the derived fractions to model the mass distribution of the di-photon candidate data.  For this we need three mass templates: $T_{\g\g}$,  $T_{\g j }$, and $T_{jj}$. We take $T_{\g\g}$ from {\sc pythia}  MC \cite{pythia} corrected for detector effects and reweighted with the K-factor derived from {\sc ResBos} \cite{resbos} to account for the (next-to-)next-to-leading order, NLO (NNLO), effects. The other two templates are taken from $\g j$ and $j j$ samples, where we relax the calorimeter isolation, EM fraction, and energy-weighted shower width requirements in the definition of a jet in order to increase statistics in these templates. We verify that relaxing the requirements do not alter the kinematics of the sample. We also correct the $\g j$ mass template for the admixture of $jj$ events. We construct the background mass spectrum assuming the functional form $N_{\g\g}\cdot (T_{\g\g}+c_1\cdot T_{jj}+c_2\cdot T_{\g j})$ where $T_{\g\g}$, $T_{\g j}$, and $T_{jj}$ are mass distributions normalized to one (see Fig.~\ref{fig:fig_1}), $c_1$ and $c_2$ are taken from the CPS fit above, and  $N_{\g\g}$ is the expected number of DDP events from the MC. For the measured luminosity, we estimate $N_{\g\g}=113\pm3.5(\text{stat})\pm24(\text{syst})$ events, which is in agreement with the $c_0 =  131\pm 22\pm7$ events derived from data. While these numbers, 113 and 131, are within the theoretical and experimental uncertainties, we choose to normalize the number of background events to the total number of events observed in the data (normalization events are counted outside of the signal region, defined as a $\pm5$~GeV window in diphoton mass centered at each hypothesized $m_{h_f}$ value). By doing so we eliminate most of the background uncertainties, e.g. luminosity, renormalization scale. 

\begin{figure}[htp]
\begin{center}
\includegraphics[scale=0.45, angle=90]{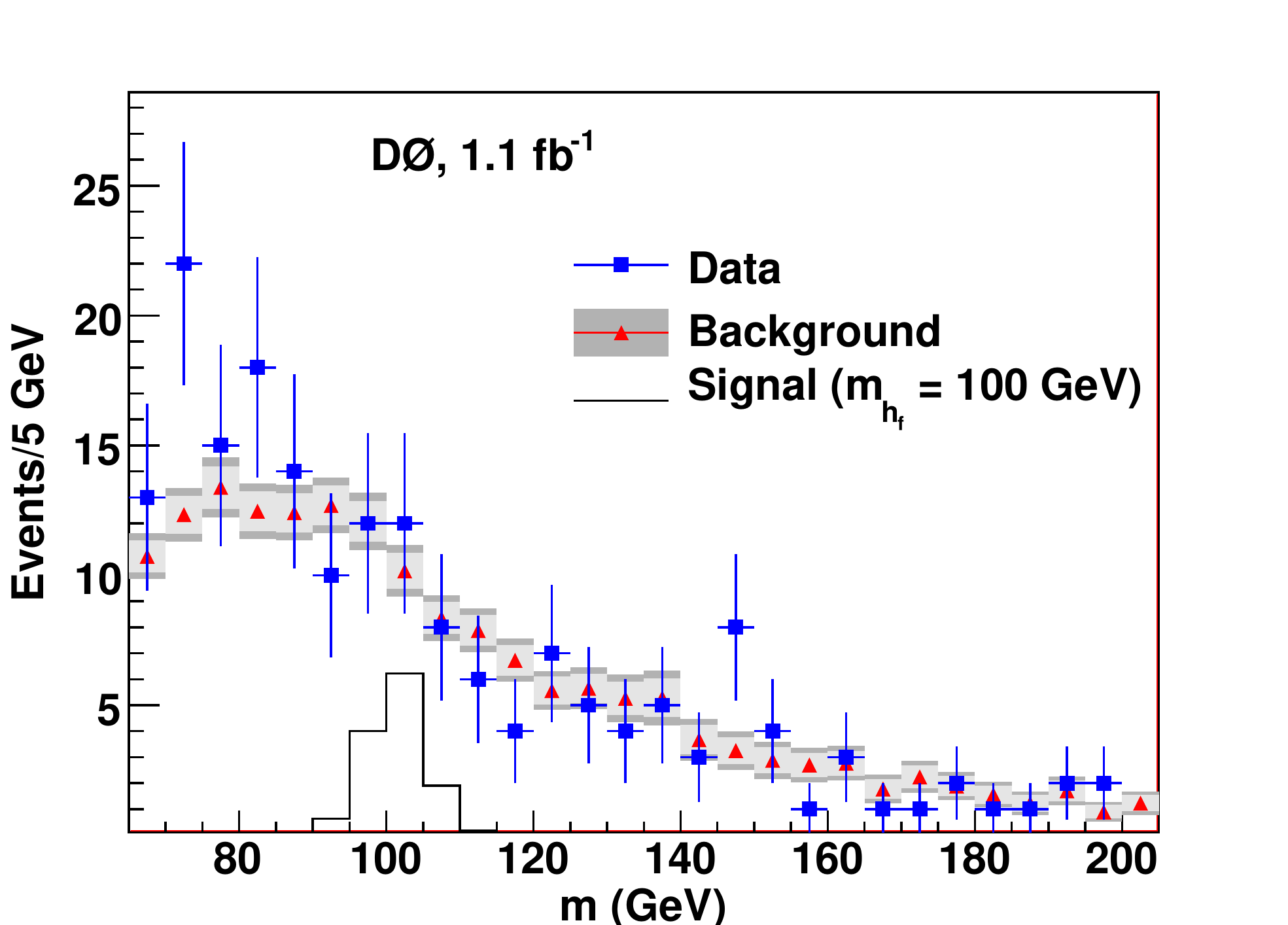}
\end{center}
\caption{Di-photon mass distribution of the data (squares) with the overlaid background prediction (triangles), and the expected signal distribution for $m_{h_f}=100$~GeV in the benchmark model. 
 \label{fig:fig_2}}
\end{figure}

Figure \ref{fig:fig_2} shows the mass distributions in data with overlaid background predictions. The shaded regions correspond to the expected background error bands. The inner band represents the statistical uncertainty of the mass templates, while the outer corresponds to the systematics due to variation in the one-dimensional $\sigma_E^{\text{CPS}}$ templates. We assign an additional 100\% uncertainty that includes any possible change in the shape of the mass templates due to the relaxed definition of a jet.
 
Signal events are generated for a range of mass points from 70 GeV to 150 GeV in 10 GeV steps. We use the {\sc pythia} event generator followed by a detailed {\sc geant}-based \cite{geant} simulation of the \Dzero detector. The signal efficiencies, $\epsilon^{\text{signal}} $, are derived from the MC. Table~\ref{tab:summary} lists signal efficiencies after correction for trigger inefficiency and scaling by the ratio of efficiencies in data and MC ($\approx95\%$ per photon) obtained from  the electron reconstruction efficiency in $Z\rightarrow e^+e^-$ events. Note that the photon requirements are chosen in such a way that the MC correctly reproduces differences between electrons and photons as confirmed in $Z\rightarrow e^+e^-\g$ events. Table~\ref{tab:summary} also shows the number of observed di-photon candidate events in data in 10 GeV mass windows  and the corresponding background estimates with associated uncertainties. The width of the mass peak is dominated by the detector resolution and varies between 2.8 GeV and 5.2 GeV. The size of the optimal mass window varies between 8 GeV and 15 GeV, but for simplicity we use a fixed value of 10~GeV. The acceptance of the mass window cuts varies between 94\% and 66\% for $m_{h_f}=70-150$ GeV. In the same table we provide the theoretical benchmark branching ratio, $B(h\rightarrow \g\g)$ \cite{benchmark_br}, and the NLO cross section, $\sigma^{\text{NLO}}_h$, for the sum of the signal processes $p\bar{p}\rightarrow VV\rightarrow h_f$ and  $p\bar{p}\rightarrow h_f V$ obtained with {\sc vv2h} and {\sc v2hv}~\cite{HIGLU}.
	
\begin{table*}
\begin{center}
%\begin{ruledtabular}{|c|c|c|c|c|c|c|c|c|}
\begin{tabular}{ccccccccc}
\hline
\hline
&&&&\multicolumn{3}{c}{ $\sigma(p\bar {p}\rightarrow h_f+X)\cdot B(h_f\rightarrow \g\g)$ (pb)}\\
%$m_{h_f}$ (GeV) &$N_{\text{data}}$&$N_{\text{bkg}}$&$\epsilon^{\text{signal}}$ (\%)&$\sigma_{\text{exp}}^{\text{limit}}\cdot B$ (pb)&$\sigma_{\text{obs}}^{\text{limit}}\cdot B$ (pb) & $\sigma_{\text{Run I}}^{\text{limit}}\cdot B$ (pb)\&$\sigma^{\text{NLO}}_{h}$ (pb)& $B(h_f\rightarrow \g\g)$\\ 
%\hline

$m_{h_f}$ (GeV) &
data&
background&
$\epsilon^{\text{signal}}$(\%)                                &
expected limit  &
observed limit   &
Run I limit   &
$\sigma^{\text{NLO}}_{h}$ (pb)                             &
$B(h_f\rightarrow \g\g)$\\ 
\hline

70&35&$24.5\pm4.6$&$ 6.9\pm0.5$   &0.15 &0.29&0.46 &1.5 &0.81       \\ 
 80&33&$27.2\pm5.0$&$ 7.9\pm0.6$  &0.14 &0.20&0.44 &1.0 &0.70       \\ 
 90&24&$27.4\pm5.4$&$ 9.8\pm0.8$  &0.11 &0.089&0.37&0.75&0.41     \\ 
 100&24&$23.7\pm4.8$&$ 10.3\pm0.8$&0.10 &0.10&0.35 &0.55&0.18    \\ 
 110&14&$17.7\pm4.4$&$ 11.2\pm0.9$&0.085&0.061&0.34&0.42&0.062 \\ 
 120&11&$13.4\pm3.7$&$ 11.3\pm0.9$&0.070&0.058&0.33&0.32&0.028    \\ 
 130&9&$11.7\pm3.3$&$ 11.2\pm0.9$ &0.065&0.053&0.33&0.25&0.019    \\ 
 140&8&$9.5\pm2.8$&$ 11.7\pm0.9$  &0.058&0.052&0.32&0.19&0.0061  \\ 
 150&12&$6.3\pm2.1$&$ 11.7\pm0.9$ &0.051&0.10&0.32 &0.15&0.0020  \\ 
 \hline
 \end{tabular}
%\end{ruledtabular}
\end{center}\caption{ Input data for limit calculation and 95\% C.L. limits on cross section times branching fraction. 
Quoted are the total uncertainties that are used in the limit calculation. \label{tab:summary}}
\end{table*}

\begin{figure}[htp]
\begin{center}
\includegraphics[scale=0.45, angle=90]{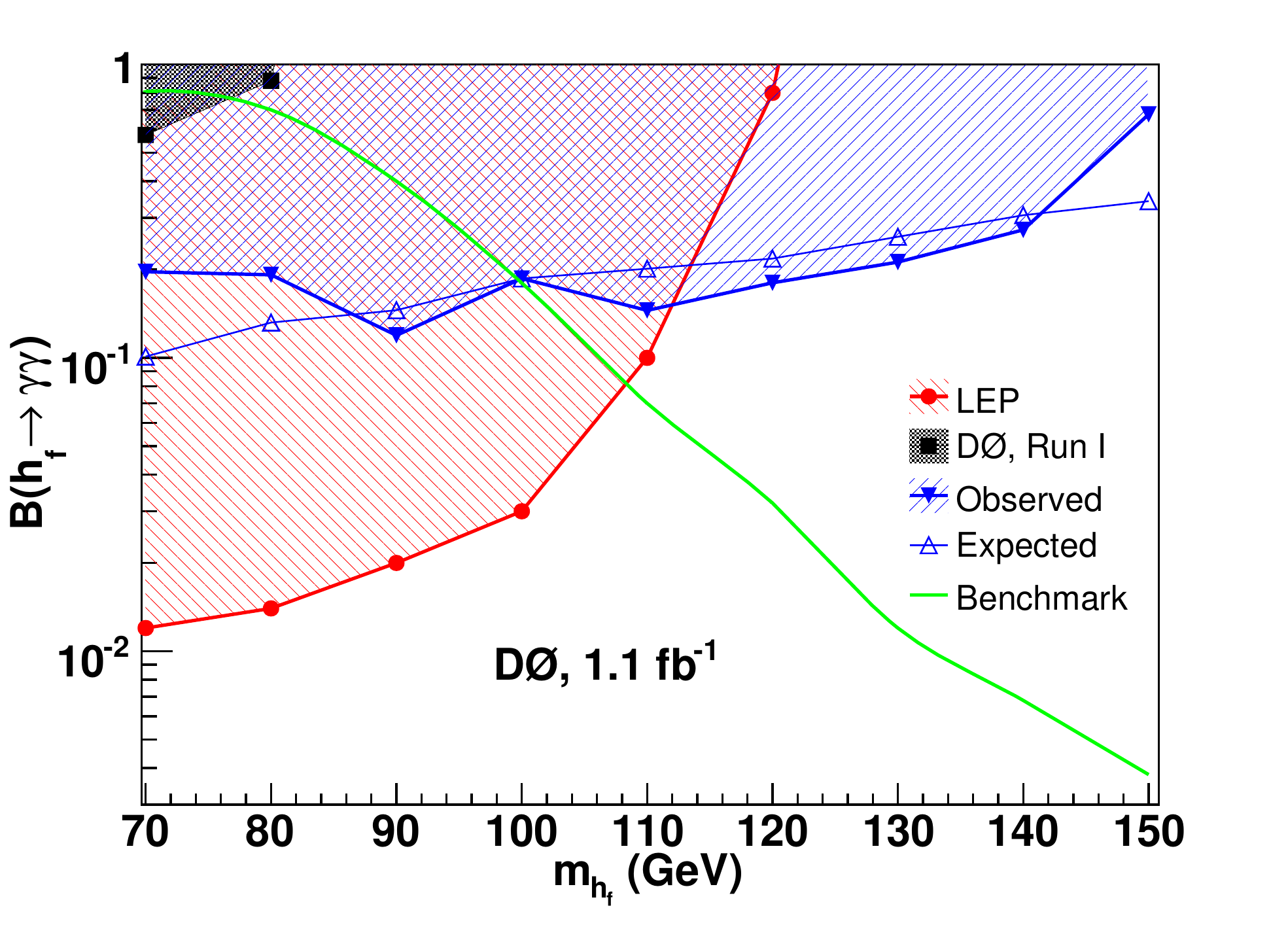}
\end{center}
\caption{$B(h_f\to\gamma\gamma)$ limits as a function of the Higgs mass. The theoretical
$B(h_f\to\gamma\gamma)$ curve 
for the benchmark model as well as the observed $B(h_f\to\gamma\gamma)$ limits from \Dzero Run I and
LEP are overlaid. The shaded regions correspond to the excluded values of the branching ratio.
\label{fig:fig_3}}
\end{figure}

We perform a counting experiment in the 10 GeV mass windows, and in the absence of an excess of di-photon events, we set
an upper limit on the product of the Higgs boson production cross section and
di-photon branching ratio $\sigma_{h_f}\cdot B(h_f\to\gamma\gamma)$ at 95\% C.L. Limits
are calculated using the modified frequentist $CL_{S}$ method \cite{cls}.
Table~\ref{tab:summary} shows the expected and observed limits. %($B\cdot \sigma_{\text{exp}}^{\text{limit}}$ and $B\cdot \sigma_{\text{obs}}^{\text{limit}}$ respectively). 
The present study excludes fermiophobic Higgs bosons of mass up to 100 GeV at the 95\% C.L. This is the most stringent limit to date at a hadron collider. In Fig.~\ref{fig:fig_3} we present our results as limits on the branching ratio in the parameter space $B(h_f\to\gamma\gamma)$ vs. $m_{h_f}$ obtained from a ratio of the above limits and $\sigma^{\text{NLO}}_{h}$. The regions above the experimental points correspond to the excluded values of the branching ratio. This study significantly improves the LEP limits at intermediate mass values, e.g. by more than a factor of four at $m_{h_f}=120$~GeV, and extends sensitivity into the region not accessible at LEP, $m_{h_f}>130$ GeV. 

% acknowledgement_paragraph_r2.tex                         2/19/08
%
We thank the staffs at Fermilab and collaborating institutions, 
and acknowledge support from the 
DOE and NSF (USA);
CEA and CNRS/IN2P3 (France);
FASI, Rosatom and RFBR (Russia);
CNPq, FAPERJ, FAPESP and FUNDUNESP (Brazil);
DAE and DST (India);
Colciencias (Colombia);
CONACyT (Mexico);
KRF and KOSEF (Korea);
CONICET and UBACyT (Argentina);
FOM (The Netherlands);
STFC (United Kingdom);
MSMT and GACR (Czech Republic);
CRC Program, CFI, NSERC and WestGrid Project (Canada);
BMBF and DFG (Germany);
SFI (Ireland);
The Swedish Research Council (Sweden);
CAS and CNSF (China);
and the
Alexander von Humboldt Foundation.

\end{document}